\input harvmac
\def \bp {{\bar \phi}}

\def \ep {\epsilon}
\def \const {{\rm const}}
\def \ha{{\textstyle{1\over 2}}}
\def \qu{{\textstyle{1\over 4}}}
\def \hai{{\textstyle{i\over 2}}}
\def \hal{{\textstyle{1\over 2}}}

\def \a {\alpha}
\def \b {\beta}

\def \p {\phi}

\def \vp {\varphi }
\def \l {\lambda}
\def \t {\theta}
\def \td {\tilde }
\def \d {\delta}
\def \bt {\bar \theta}

\def \bD {\bar D}
\def \LL {{\bf L}}
\def \fo {{\textstyle{1\over 4}}}

\def \da {{\dot \a}}
\def \W {{\cal W}}
\def \WW {{\bf W}}

\def \inv {^{-1}}
\def \ov {\over }
\def \four{{\textstyle{1\over 4}}}
\def \fourth{{{1\over 4}}}

\def \L {\Lambda}
\def \LLL {{\bf \Lambda}}
\def \A {{\cal A}}
\def \E {{\cal E}}
\def \B {{\cal B}}
\def \cL {{\cal L}}\def \cA {{\cal A}}
\def \tF {\tilde F}
\def \F {{F^*}}
\def \cF {{\cal F}}

\def \G {{\cal G}}
\def \exw {\langle {\bf W} \rangle}
\def \exwb {\langle \bar{\bf W} \rangle}
\def \pa {\partial}
\def \ada {{\a\da}}
\def\ie {{\it i.e.}}
\def\eg {{\it e.g.}}
\def \lr { \lref}
\def\np {{ Nucl. Phys. }}
\def \pl {{ Phys. Lett. }}
\def \mpl {{ Mod. Phys. Lett. }}
\def \prl {{ Phys. Rev. Lett. }}
\def \pr {{ Phys. Rev. }}
\def \ap {{ Ann. Phys. }}

\baselineskip8pt
\Title{
\vbox
{\baselineskip 6pt{\hbox{ITP-SB-98-68}}
{\hbox{Imperial/TP/98-99/013
}}{\hbox{hep-th/9811232}} {\hbox{
}}} }
{\vbox{\centerline {Partial breaking of global $D=4$ supersymmetry,
}
\vskip4pt
\centerline { constrained superfields, and 3-brane actions }
}}
\vskip -20 true pt
\medskip
\medskip
\centerline { { M. Ro\v cek\footnote{$^*$} {e-mail address:
rocek@insti.physics.sunysb.edu
} }}
\smallskip \smallskip
\centerline{\it Institute of Theoretical Physics,
State University of New York }
\smallskip
\centerline{\it Stony Brook, NY 11794-3840, USA }
\medskip
\centerline {and}
\medskip
\centerline{ A.A. Tseytlin\footnote{$^{\star}$}{\baselineskip8pt
e-mail address: tseytlin@ic.ac.uk}\footnote{$^{\dagger}$}{\baselineskip8pt
Also at Lebedev Physics
Institute, Moscow.} }
\smallskip\smallskip
\centerline {\it Theoretical Physics Group, Blackett Laboratory}
\smallskip
\centerline {\it Imperial College, London SW7 2BZ, U.K. }
\bigskip
\centerline {\bf Abstract}
\medskip
\baselineskip10pt
\noindent
We show that the connection between
partial breaking of supersymmetry and nonlinear actions
is not accidental and has to do with constraints that
lead directly to nonlinear actions of the Born-Infeld type.
We develop a constrained superfield approach that
gives a universal way of deriving and using these actions.
In particular, we find the manifestly supersymmetric form of
the action of the 3-brane in 6-dimensional space in terms of
$N=1$ superfields by using the tensor multiplet as a tool.
We explain the relation between the Born-Infeld action and
the model of partial $N=2$ supersymmetry breaking by
a dual D-term. We represent the Born-Infeld action in a novel form
quadratic in the gauge field strengths by introducing two
auxiliary complex scalar fields; this makes duality covariance and
the connection with the $N=1$ supersymmetric extension
of the action very transparent. We also suggest a general procedure
for deriving manifestly duality symmetric actions,
explaining in a systematic way relations between previously
discussed Lorentz-covariant and noncovariant actions.
\medskip
\Date {April 1997 }
\noblackbox
\baselineskip 16pt plus 2pt minus 2pt
\lr\bg{ J. Bagger and A. Galperin, {\it Matter couplings in
partially broken extended supersymmetry},
\pl B336 (1994) 25, hep-th/9406217.}
\lr\bgg{ J. Bagger and A. Galperin,
{\it A New Goldstone multiplet for partially broken supersymmetry},
\pr D55 (1997) 1091, hep-th/9608177.}
\lref\pol{J. Polchinski,
{\it Dirichlet Branes and Ramond-Ramond charges},
\prl 75 (1995) 4724,
hep-th/9510017.}
\lr \ft {E.S. Fradkin and A.A. Tseytlin,
{\it Non-linear electrodynamics from quantized strings},
\pl B163 (1985) 123.}
\lr\bag{J. Bagger,
{\it Partial breaking of extended supersymmetry},
Nucl. Phys. Proc. Suppl. 52A (1997) 362,
hep-th/9610022.}
\lr \pap {H. Partouche and B. Pioline,
{\it Partial spontaneous breaking of global supersymmetry},
Nucl. Phys. Proc. Suppl. 56B (1997) 322,
hep-th/9702115. }
\lr \cecf{S. Cecotti and S. Ferrara,
{\it Supersymmetric Born-Infeld Lagrangians},
\pl B187 (1987) 335. }
\lr\bagw{ J. Bagger and J. Wess, {\it Partial breaking of extended
supersymmetry},
\pl B138 (1984) 105. }
\lr\kle{I.R. Klebanov, {\it
World volume approach to absorption by nondilatonic branes},
Nucl. Phys. B496 (1997) 231,
hep-th/9702076; S.S. Gubser, I.R. Klebanov
and A.A. Tseytlin, {\it String theory and classical absorption by
three-branes},
Nucl. Phys. B499 (1997) 217,
hep-th/9703040. }
\lr \dva {G. Dvali and M. Shifman,
{\it Dynamical compactification as a mechanism of spontaneous supersymmetry
breaking},
Nucl. Phys. B504 (1997) 127,
hep-th/9611213.}
\lref \man{M. de Groot and P. Mansfield,
{\it The Born-Infeld Lagrangian via string field theory},
Phys. Lett. B231 (1989) 245. }
\lr\iva{E. Ivanov and B. Zupnik, {\it Modified N=2 supersymmetry
and Fayet-Illiopoulos terms},
hep-th/9710236.}
\lr \ket{S.V. Ketov,
{\it A manifestly N=2 supersymmetric Born-Infeld action},
hep-th/9809121.}
\lref\HS{G.~Horowitz and A.~Strominger,
{\it Black strings and p-branes}, \np B360 (1991) 197.}
\lr\book{ S.J. Gates, M. Grisaru, M. Ro\v cek and W. Siegel,
{\it Superspace} (Benjamin/Cummings, 1983). }
\lr\nils{M.~Cederwall, A.~von~Gussich, B.E.W.~Nilsson, and A.~Westerberg,
{\it The Dirichlet super three-brane in ten-dimensional type IIB supergravity},
Nucl. Phys. B490 (1997) 163, hep-th/9610148.}
\lr\dul{M.J. Duff and J.X. Lu,
{\it The selfdual type IIB superthreebrane},
\pl B273 (1991) 409. }
\lr\lei{R.G. Leigh, {\it
Dirac-Born-Infeld action from the
Dirichlet sigma model},
\mpl A4 (1989) 2767.}
\lr\jhs{M. Aganagic, C. Popescu and J.H. Schwarz,
{\it D-brane actions with local kappa symmetry},
Phys. Lett. B393 (1997) 311, hep-th/9610249;
{\it Gauge invariant and gauge fixed D-brane actions},
Nucl. Phys. B495 (1997) 99, hep-th/9612080.}
\lr\bert{E. Bergshoeff and P.K. Townsend, {\it Super D-branes},
Nucl. Phys. B490 (1997) 145,
hep-th/9611173.}
\lr\tse{ A.A. Tseytlin,
{\it Selfduality of Born-Infeld action and Dirichlet three-brane
of type IIB superstring theory},
\np B469 (1996) 51, hep-th/9602064.}
\lr\gib{ G.W. Gibbons and D.A. Rasheed,
{\it Electric - magnetic duality rotations in nonlinear electrodynamics},
Nucl. Phys. B454 (1995) 185,
hep-th/9506035; M.K. Gaillard, B. Zumino,
{\it Nonlinear electromagnetic selfduality and Legendre transformations},
hep-th/9712103.}
\lr \hup {J. Hughes and J. Polchinski,
{\it Partially broken global supersymmetry and the superstring},
\np B278 (1986) 147.}
\lr \hulp {J. Hughes, J. Liu and J. Polchinski,
{\it Supermembranes},
\pl B180 (1986) 370.}
\lr \apt{I. Antoniadis, H. Partouche and T.R. Taylor,
{\it Spontaneous breaking of N=2 global supersymmetry},
\pl B372 (1996) 83,
hep-th/9512006.}
\lr \fgp{S. Ferrara, L. Girardello and M. Porrati,
{\it Spontaneous breaking of N=2 to N=1 in rigid and local supersymmetric
theories},
\pl B376 (1996) 275,
hep-th/9512180.}
\lr \por{M. Porrati,
{\it Spontaneous breaking of extended supersymmetry in global and local
theories },
Nucl. Phys. Proc. Suppl. 55B (1997) 240,
hep-th/9609073.}
\lr \des{S. Deser and R. Puzalowski, {\it Supersymmetric nonpolynomial
vector multiplets and causal propagation},
J. Phys. A13 (1980) 2501.}
\lr\gaz{M.K. Gaillard, B. Zumino,
{\it Duality rotations for interacting fields},
Nucl. Phys. B193 (1981) 221.}
\lr \town{ P.K. Townsend,
{\it D-branes from M-branes},
\pl B373 (1996) 68, hep-th/9512062;
C. Schmidhuber, \np B467 (1996) 146, {\it D-brane actions},
hep-th/9601003. }
\lr \ftt {E.S. Fradkin and A.A. Tseytlin,
{\it Quantum equivalence of dual field theories},
\ap 162 (1985) 31.}
\lr \tsd{ A.A. Tseytlin,
{\it Duality symmetric formulation of string world-sheet
dynamics},
\pl B242 (1990) 163;
{\it Duality symmetric closed string theory and interacting chiral scalars},
\np B350 (1991) 395.}
\lr\dese{ S. Deser and C. Teitelboim, {\it Duality transformations of
abelian and non-abelian gauge
fields}, \pr D13 (1976) 1592.}
\lr \ss {J.H. Schwarz and A. Sen, {\it Duality symmetries of 4-D heterotic
strings},
\pl B312 (1993) 105, hep-th/9305185. }
\lr \sss {J.H. Schwarz and A. Sen,
{\it Duality symmetric actions},
\np B411 (1994) 35, hep-th/9304154.}
\lr \canon {A. Giveon, E. Rabinovici and G. Veneziano, {\it Duality in
string background space},
\np B322 (1989) 167;
K. Meissner and G. Veneziano, {\it Symmetries of cosmological superstring
vacua},
\pl B267 (1991) 33;
E. \' Alvarez, L. \' Alvarez-Gaum\' e and Y. Lozano,
{\it A Canonical approach to duality transformations},
\pl B336 (1994) 183,
hep-th/9406206.}
\lr \bra { A. Khodeir and N. Pantoja, {\it Covariant duality symmetric
actions},
\pr D53 (1996) 5974,
hep-th/9411235.}
\lr \pas {P. Pasti, D. Sorokin and M. Tonin, 
{\it Note on manifest Lorentz and general coordinate invariance in
duality symmetric models}, 
\pl B352 (1995) 59,
hep-th/9503182;  {\it Duality symmetric actions with manifest space-time symmetries}, 
 Phys. Rev. D52 (1995) 4277, hep-th/9506109;
hep-th/9607171. }
\lr \pasi {P. Pasti, D. Sorokin and M. Tonin,
{\it On Lorentz invariant actions for chiral p forms},
Phys. Rev. D55 (1997) 4332,
hep-th/9611100.}
\lr \busc {T.H. Buscher, {\it A symmetry of the string background
field equations},
\pl B194 (1987) 59; {\it Path integral derivation
of quantum duality in nonlinear sigma models},
\pl B201 (1988) 466.}
\lr\rocver{ M. Ro\v cek and E. Verlinde, {\it Duality, quotients, and
currents},
\np B373 (1992) 630, hep-th/9110053. }
\lr \wit{E. Witten, {\it On S duality in Abelian gauge theory},
hep-th/9505186; E. Verlinde, {\it Global aspects of electric - magnetic
duality},
Nucl. Phys. B455 (1995)
211, hep-th/9506011; J.L.F. Barbon,
{\it Generalized Abelian S duality and coset constructions},
Nucl. Phys. B452 (1995) 313,
hep-th/9506137. }
\lr \jac {R. Floreanini and R. Jackiw,
{\it Selfdual fields as charge density solitons},
\prl 59 (1987) 1873.}
\lr\teit{ M. Henneaux and C. Teitelboim,
{\it Dynamics of chiral (selfdual) p-forms},
\pl B206 (1988) 650.}
\lr\oth{ M. Aganagic, J. Park, C. Popescu and J.H. Schwarz,
{\it World volume action of the M theory five-brane},
Nucl. Phys. B496 (1997) 191,
hep-th/9701166;
I. Bandos, K. Lechner, A. Nurmagambetov, P. Pasti, D. Sorokin and
M. Tonin, 
{\it Covariant Action for the Super-Five-Brane of M-Theory}, 
Phys. Rev. Lett. 78 (1997) 4332, hep-th/9701149; 
  {\it On the equivalence of different
formulations of the M theory five-brane},
Phys. Lett. B408 (1997) 135,
hep-th/9703127.}
\lr \martin{M. Ro\v cek, {\it Linearizing the Volkov-Akulov model},
\prl 41 (1978) 451.}
\lr \tst { A.A. Tseytlin, {\it On non-abelian generalization of
Born-Infeld action in string theory},
Nucl. Phys. B501 (1997) 41, hep-th/9701125. }
\lr \bggg{ J.~Bagger and A.~Galperin, {\it The Tensor Goldstone
multiplet for partially broken supersymmetry},
Phys. Lett. B412 (1997) 296, hep-th/9707061.}
\lr \some{P.~Fayet, {\it Spontaneous generation of massive central 
charges in extended supersymmetric theories},
 Nucl. Phys. B149 (1979) 137;
M.T.~Grisaru, W.~Siegel, and M.~Ro\v cek, {\it Improved methods for
supergraphs},  Nucl. Phys. B159 (1979) 429. }
\lr \vec{R. Grimm, M. Sohnius and J. Wess,
{\it Extended supersymmetry and gauge theories},
\np B133 (1978) 275. }
\lr \tens{J.\ Wess, {\it Supersymmetry and internal symmetry},
Acta Phys. Austriaca 41 (1975) 409;
W.\ Siegel, {\it Off-shell central charges}, Nucl.Phys.B173 (1980) 51.}
\lr \sw{N. Seiberg and E. Witten, {\it
Electric - magnetic duality, monopole condensation, and
confinement in N=2 supersymmetric Yang-Mills theory},
\np B426 (1994) 19, hep-th/9407087. }
\lr \gpr{F. Gonzalez-Rey, I.Y. Park, and M. Ro\v cek,
{\it On dual 3-brane actions with partially broken N=2 supersymmetry},
hep-th/9811130.}
\lr \lamb{N.D. Lambert  and  P.C. West, 
{\it Brane dynamics and four-dimensional quantum field theory}, 
hep-th/9811177.}

\lr\ivano {S. Bellucci, E. Ivanov and S. Krivonos, 
{\it Partial breaking N=4 to N=2: hypermultiplet as a
Goldstone superfield},  hep-th/9809190.}

\lr\gir{
P. Fre, L. Girardello,
 I. Pesando     and  M. 
    Trigiante, {\it Spontaneous  ${N=2 \to  N=1}$  local supersymmetry  
    breaking   with surviving    compact gauge group,}
    Nucl. Phys. B493  (1997) 231, 
   hep-th/9607032.  } 

\lr\GIR{S. Cecotti, L. Girardello and M. Porrati,
{\it ``Some features of susy breaking in $N=2$ supergravity},  
Phys.
Lett. B151 (1985) 367;
{\it ``An exceptional $N=2$ supergravity 
with flat potential and partial superhiggs",} Phys. Lett. 
 B168 (1986) 83.}

\newsec{Introduction}
Partial breaking of global $N=2$ supersymmetry has been
discussed from various points of view \refs{\bagw,\hup,
\hulp,\bg,\apt,\fgp,\gir,\bgg,\dva, \iva} (see also reviews in
\refs{\bag,\por,\pap}).\foot{Partial breaking 
of $N=2$ to $N=1$ supersymmetry in the context of a  supergravity 
theory (without negative norm states) 
was 
originally discussed in  \GIR.}
 Several ways to construct models
with partial supersymmetry breaking in various dimensions are known. One
is a `solitonic' realisation of partial supersymmetry breaking that
uses a BPS soliton background in a higher-dimensional
supersymmetric theory. Such a solution breaks some of translational
invariances and thus some of supersymmetries. As a result, a
lower-dimensional action for the corresponding collective coordinates
has part of the supersymmetry realised linearly and part nonlinearly
\refs{\hup,\hulp}.\foot{Starting with a higher-dimensional  vector gauge theory
leads only to scalar multiplet actions for the collective coordinates.
To get a massless vector ($D\geq 4$) or tensor ($D\geq 6$) multiplet one
must consider solitons in a theory containing higher tensors
such as a  tensor multiplet theory in $D=6$ or  supergravity.}

Another approach is to start with a generalised (nonrenormalisable)
$N=2$, $D=4$ vector multiplet action and to add
a dual (magnetic) D-term \apt. Here
one works directly in four dimensions and the vacuum is translationally
invariant but because of the nonlinear structure of the action,
the `magnetic' D-term can spontaneously break
$N=2$ supersymmetry to $N=1$. This model can be also derived as a
special limit of an $N=2$ supergravity model \fgp.

One of the aims of the present paper is to achieve a better understanding
of how to construct nonlinear actions of theories with partially broken
supersymmetry in a closed, manifestly supersymmetric way. In \martin, one
of us showed how to describe the goldstino of spontaneously broken
$N=1$ supersymmetry in terms of a constrained chiral superfield. In this
paper, we use the same approach to construct the $N=1$ superfield
goldstone multiplets of partially broken $N=2$ supersymmetry. One example
discussed previously in the literature is the $D=4, N=1$ supersymmetric
Born-Infeld (BI) action \refs{\des,\cecf} which can be interpreted \bgg\
as the unique action corresponding to the situation when $N=2$
supersymmetry is broken down to $N=1$ in such a way that the vector
multiplet remains massless (\ie, is the Goldstone multiplet). Our method
gives a systematic derivation of the BI action of \bgg , as well
as a supersymmetric membrane action in terms of a tensor multiplet
described by a real linear superfield.\foot{While we were in the process
of writing up our results, this action appeared in \bggg , but without a
systematic derivation.}

We also explain how the $N=1$ super Born-Infeld
action emerges from the model of \apt\ when one decouples (`integrates
out') the massive chiral multiplet, and address the case of $N=2\to 1$
supersymmetry breaking with the scalar multiplet remaining massless. This
case is related to the model of \hulp\ and was previously discussed in
\bg\ where, however, the full nonlinear form of the corresponding $N=1$
chiral multiplet action was not determined. This action should be the
manifestly $N=1, D=4$ supersymmetric form of the Nambu-type action
for
the 3-brane in six dimensions of ref. \hulp\ in the static gauge.
We propose a class of chiral multiplet actions.
We discuss the
superfield analog of the
$D=4$ scalar-tensor duality in the case of the nonlinear $D=4$ Nambu action,
explaining that one of the chiral multiplet actions should also have a
second hidden supersymmetry.

Part of the motivation behind this work is to try to determine the
complete $N=1,D=4$ superfield form of the static-gauge action for the D3-brane
soliton of the type IIB supergravity (string theory)
\refs{\HS,\dul,\pol}. The bosonic part of this action is a `hybrid' of
the $D=4$ BI and Nambu actions \lei\ (closely related -- by
T-duality -- to the $D=10$
BI action \ft)
\eqn\one{ S= - \int d^4 x \sqrt{- \det( \eta_{ab}
+ \del_a X^n \del_b X^n + F_{ab})} \ , \ \ \ \ \ \ \ \ \ n=1,...,6 \ . }
Here $a,b=0,1,2,3$ are the world-volume indices and $X^n$
are collective
coordinates corresponding to the `transverse' motion of a 3-brane in $D=10$
space (we set the string tension to 1, $T\inv =2\pi\a'=1$,
and ignore the
overall factor of the D3-brane tension).
The component form of the supersymmetric
extension of this action with $4$ linearly realised and $4$ nonlinearly
realised global supersymmetries was found in
\refs{\nils,\jhs,\bert}. The leading term in this action is the $N=4,
D=4$ super-Maxwell action. It would be interesting for several reasons
(the quantum properties of D3-branes
and their comparison with supergravity \kle,
possible hints about a non-abelian
generalisation, etc.) to have a manifestly supersymmetric formulation
of this action in terms of one vector and three chiral $N=1,D=4$ superfields.
That would be a Born-Infeld generalisation of the corresponding
unconstrained superfield form of the $N=4,D=4$ super-Maxwell action \some.

Since the action \one\ contains both the vector {\it and} the scalars
the knowledge of the $N=1$ superfield form of the $D=4$ BI action for a
single vector field \refs{\des,\cecf,\bgg} is by far not sufficient to
determine its superfield analogue. As a step towards understanding how to
combine the vector and scalar dependencies one may try first to determine
the superfield form of the
action for the 3-brane of ref.~\hulp, which does not
contain a vector field. The model considered in \hulp\ was the $N=1,D=6$
supersymmetric theory of Maxwell multiplet coupled to two (charged)
scalar multiplets. The 3-brane solution in this theory (the direct BPS
analogue of the Abrikosov-Nielsen-Olesen string in $D=4$) breaks
translational invariance in 2 of 5 spatial directions and thus breaks
half of $N=1,D=6$ or, equivalently, $N=2,D=4$,
supersymmetry. The resulting
static-gauge action for the 3-brane collective coordinates has
the following bosonic part
\eqn\two{
S'= - \int d^4 x
\sqrt{- \det( \eta_{ab} + \del_a X^n \del_b X^n)}\ , \ \ \ \ \ \ \ \
n=1,2 \ .}
We thus seek a superspace extension of \two\ with manifest $N=1,D=4$
supersymmetry as well as the broken `half' of $N=2$ supersymmetry realised
nonlinearly.

 Below we shall determine the supersymmetric form of \two\
using two
different approaches. One is a direct $N=1$ chiral multiplet
construction that has the right bosonic part matching \two. To
demonstrate that this action has also hidden $N=2$ supersymmetry we
shall use the $D=4$ duality transformation that `rotates' one of the two
scalars in \two\ into an antisymmetric 2-tensor. It turns out to be
simpler to construct the superfield action with partially broken $N=2$
supersymmetry for a {\it tensor} multiplet rather than for
a chiral multiplet.\foot{This was previously considered in \bg\ using a different
approach.  The first two terms in the expansion of the
supersymmetric analog of \two\ are related to the action found in \bg\
by a field redefinition which eliminates terms in the action of \bg\
which are not invariant under $X^n\to X^n + a^n, \ a^n= \const.$ Our
tensor multiplet results were also found in \bggg, where a dual chiral
superfield action that agrees with our general form is also proposed; see
also \gpr.}

In section 2 we discuss constrained superfields and how they may be
used to express linearly transforming fields in terms of Goldstone
fields. In particular, we apply this to partial supersymmetry breaking in
the vector and tensor multiplets, and find the relation to the other
examples of partial supersymmetry breaking described above.
In section 3 we apply the lessons learned from the supersymmetric
case to rewrite the Born-Infeld action in various forms that make duality
transparent. We also consider scalar-tensor duality in the
Born-Infeld-Nambu actions.
In section 4 we give our conclusions and mention some open problems.
In the Appendix we discuss in detail a manifestly duality invariant
formalism for general systems,
explaining relations between previously
considered  Lorentz-covariant and noncovariant actions.
\newsec{Constrained superfields and partial supersymmetry breaking}
Spontaneously broken symmetries give nonlinear realizations of the
broken symmetry group. A traditional way to find such realizations is
to begin with a linear representation and impose a nonlinear
constraint (\eg, spontaneously broken rotational symmetry may
be nonlinearly realized on a vector constrained to lie on the surface
of a sphere). In \martin, it was shown that an $N=1$ chiral superfield
obeying the nonlinear constraints\foot{We mostly use the conventions of
\book ; in particular, $D^2=\ha D^\a D_\a$,\ and the $2\times 2$
charge conjugation matrix $C_{\a\b}$ is  $i\pmatrix{ 0 & -1 \cr
1 & 0 \cr}.$ However,  the matrix 
$C_{ab}$  contracting  the indices $a,b=1,2$  which label
the two  supersymetries is defined without a factor of $i$.}
\eqn\nlcon{(i) \ \ \phi=\bar D^2 (\phi\bar\phi)\ ,\ \ \ \ \ \phi^2=0\ ; \ \
\ \ \ \
\ \ (ii) \ \ \langle D^2\phi\rangle=1\ ,}
can be expressed in terms of a single fermionic field: the goldstino
for the broken $N=1$ supersymmetry. The two constraints can be understood
as follows: the constraint on $\langle D^2\phi\rangle$ implies that
supersymmetry is spontaneously broken,\foot{We shall use
$ \langle ...\rangle $ to indicate that a (super)field has a classical vacuum
expectation value.}
whereas the nonlinear constraint
removes the `radial' degrees of freedom to leave only the goldstone
field.  We shall follow the same approach to find $N=1$ superfield descriptions
of the goldstone multiplets that arise when $N=2$ supersymmetry is
partially broken to $N=1$. We consider different $N=2$ superfields
and find different $N=1$ goldstone multiplets. The two examples that
we have worked out below are the two simplest irreducible $N=2$ multiplets:
the vector and tensor multiplets.
\subsec{The Vector Multiplet}
The $N=2$ vector multiplet is described by a constrained chiral field
strength $\WW(x,\t_1,\t_2)$ that obeys the Bianchi identity \vec\
($a,b=1,2$)
\eqn\vbi{D^2_{ab}\WW=C_{ac}C_{bd}\bar D^{2cd}\bar \WW\ .}
It is convenient to define $D\equiv D_1 ,\ Q\equiv D_2$ and to rewrite
\vbi\ as:
\eqn\vbiq{D^2\WW=\bar Q^2\bar\WW\ ,\ \ \ \ \ DQ\WW=-\bar D\bar Q \bar\WW\ .}
We break $N=2$ supersymmetry to $N=1$ by assuming that $\WW$ has a
Lorentz-invariant condensate $\exw$:
\eqn\ntwocon{\exw= -\t^2_2\ ,\ \ \ \ \ \langle Q^2\WW \rangle = 1\ ,\ \ \ \ \
D\exw= 0\ . }
Here we set the scale of the supersymmetry breaking to $1$. We reduce the
field content to a single $N=1$ superfield by imposing
\eqn\nww{\W^2=0\ ,} where $\WW\equiv\W + \exw$.
Then $Q^2(\W+\exw)=\bar D^2 (\bar\W +
\exwb)$ implies
\eqn\nltr{Q^2\W =\bar D^2 \bar\W - 1\ ,}
and the constraint $\W^2=0$ implies
\eqn\qqww{0=\ha Q^2\W^2=\W(\bar D^2 \bar\W - 1)+
\ha Q^\a\W Q_\a\W\ .}
Projecting to $N=1$ superspace by setting $\t_2=0$ and defining the
$N=1$ superfields
\eqn\nldef{\p\equiv\W|_{\t_2=0}\ ,\ \ \ \ \ \ \
W_\a\equiv-Q_\a\W|_{\t_2=0}\ ,}
we find
\eqn\vecon{\p=\p\bar D^2\bp+\ha W^\a W_\a\ .}
This is precisely the constraint of \bgg. However, now we have an
interpretation of the chiral object that they mysteriously introduced:
it is the chiral superpartner of the vector multiplet in the $N=1$
superspace description of the $N=2$ vector multiplet. Note the
close analogy between \nlcon\ and \ntwocon , \nww , \vecon . Note also
that \qqww\ and \ntwocon\ actually imply \nww .
The $N=2$ supersymmetry transformation laws in superspace
follow from the constraints and definitions of the $N=1$ superfield
components, and are given by
\eqn\trans{\d_2\p\equiv(\eta^\a Q_\a+\bar\eta^\da \bar Q_\da)\W|_{\t_2=0}
=-\eta^\a W_\a\ ,\ \ \ \ \ \d_2W_\a=\eta_\a(\bar D^2\bar\p - 1)-i\bar\eta^\da
\partial_{\a\da}\p\ ,}
which, up to conventions, agrees with \bgg .
\subsec{Actions for the Vector Multiplet}
Because of the constraints \ntwocon , \nww , \nltr , there are many
equivalent forms that all give the same action. In particular, a broad
class of actions is proportional to the $N=2$ Fayet-Iliopoulos term:
$$L= \int d^2 \t_1 d^2 \t_2\ \cF (\WW)
\qquad\qquad\qquad\qquad\qquad\qquad\qquad\qquad\qquad\qquad$$
$$ = \int d^2 \t_1 d^2
\t_2 \left[\cF(0) + \cF'(0)(\exw+\W) +\ha\cF''(0) (\exw+\W)^2\right]$$
\eqn\acts{ = \cF''(0) \int d^2 \t_1\ \p \ . \ \ \
\qquad\qquad\qquad\qquad\qquad\qquad\qquad\qquad\qquad}
We can also write first-order actions where we
impose the constraints by a chiral restricted $N=2$ Lagrange multiplier
$\LLL$
\eqn\laaa{
D^2\LLL=\bar Q^2\bar\LLL\ , \ \ \ \ \ \ \ DQ\LLL=-\bar D\bar Q \bar\LLL \ , }
namely,
\eqn\uuu{ L_1=\left(\int d^2 \t_1 d^2 \t_2\ \hai \LLL \W^2\ +\ \int d^2
\t_1\ \W
\ \right) + h.c. \ , }
where we have scaled $\cF''(0)\to 1$.
In $N=1$ superspace, this reduces to ($\LLL \to \L,\chi_\a$)
$$L_1 = i\int d^2\t d^2\bt \left[\ha (\bar \L \p^2 - \L\bp^2) + (\L - \bar
\L) \p\bp \right] $$
\eqn\yyy{
+\ \left( \int d^2 \t \left[i( - \L \p + \chi^\a W_\a \p + \ha\L W^\a W_\a)
+ \p\right]\
+\ h.c. \right) \ . }
Integrating over the $N=1$ Lagrange multiplier superfields
$\L$ and $\chi_\a$ we get the constraint \qqww\
(after using \nww, which follows from the boundary condition
\ntwocon).

Another action that gives the same constraints and final $N=1$
Born-Infeld action is the standard free $N=2$ vector action,
\ie, the free $N=1$ action for the vector ($V$) and chiral ($\p$)
superfields, plus a constraint term with a chiral $N=1$
superfield Lagrange multiplier $\L$:
\eqn\iui{ S= \int d^4 x \bigg( \int d^2 \t \bigg[ ( \ha W^\a W_\a + \p \bD^2
\bp ) + i \L (\ha W^\a W_\a + \p \bD^2 \bp - \p) \bigg] + h.c. \bigg) \ , }
or, equivalently, after shifting $\L \to \L + i$,
\eqn\cti{ S = \int d^4 x \bigg[ \int d^2 \t \bigg(
i\L \big[\ha W^\a W_\a + \p \bD^2 \bp - \p\big] \ +\ \p\ \bigg)
+\ h.c. \bigg] \ . }
The explicit solution of the constraint \vecon\ is \bgg
\eqn\solu{\p (W,\bar W) = \ha W^\a W_\a
+\ha \bD^2 \bigg[ {W^\a W_\a \bar W^\da \bar W_\da
\ov 1 - \ha A + \sqrt { 1 - A + \qu (D^2 (W^\a W_\a ) -
\bD^2 (\bar W^\da \bar W_\da ))^2 } } \bigg] \ , }
where
$$A\equiv D^2 (W^\a W_\a ) + \bD^2 (\bar W^\da \bar W_\da )\ . $$
Note that adding a
D-term for $V$ is not allowed here -- this will break $N=1$
supersymmetry as there are non-minimal terms in the action (the
dependence of the bosonic part of the action on the auxiliary field
D can be found
by shifting $F^2 \to F^2 - 2 {\rm  D}^2, \ F\td F \to F\td F$ in the
bosonic BI action, see below).

The resulting action \acts\ is thus simply
\eqn\pti{ S= \int d^4 x \bigg[ \int d^2 \t \ \p(W,\bar W) +
h.c. \bigg] \ , } \ie, $\p(W,\bar W)$ is nothing but the $N=1$
supersymmetric BI action of \cecf. We thus arrive at the same conclusion
as \bgg: {\it the requirement of partially broken $N=2$ supersymmetry
uniquely fixes the action for the vector multiplet to be the
supersymmetric Born-Infeld action}.

The Lagrange multiplier form of the supersymmetric BI action \cti\
dramatically simplifies the proof \bgg\ of the {\it duality covariance}
of
the $N=1$ BI action \pti,\solu\ (this is also true in the bosonic BI action
case, see below). We relax the reality constraint on the chiral
superfield $W_\a$ and add the term with the dual
field strength $\td W_\a$ as the Lagrange multiplier:
\eqn\ctit{ \tilde S= \int d^4 x \bigg( \int d^2 \t \bigg[
\ i\L (\ha W^\a W_\a + \p \bD^2 \bp - \p) + \p - i \td W^\a W_\a \bigg] \ + \
h.c.
\bigg) \ . }
Integrating out $W_\a$ gives back the same supersymmetric BI action
\cti\ with
\eqn\dal{W\to \td W \ , \ \ \ \ \ \ \
\L\to - {\L\inv} \ , \ \ \ \ \ \ \ \ \ \ \p\to -i \L \p \ , }
This is the direct superfield analogue of the transformation
found in the bosonic case (see below).
In $N=2$ notation, this is precisely the duality transformation
of \sw\ with $\L$
playing the role of $\tau =\cF''$.
\subsec{Born-Infeld Action From the `Dual {\rm D}-term' Model }
In \apt, partial supersymmetry breaking was induced by `magnetic'
Fayet-Iliopoulos (FI) terms. Consider the action
\eqn\fif{
S=-Im\int d^4x\left( \int d^2 \t \left[\ha \cF''(\p) W^\a W_\a + m \cF' (\p) +
(e-i\xi)\p \right] + \int d^2 \t d^2 \bt \ \cF' (\p) \bp \right) \ . }
We find it convenient to introduce the $N=2$ FI terms in the $N=1$
superpotential, and {\it not\ } as D-terms. As in \apt, we choose the
magnetic FI coefficient $m$ to be real; however, since we have rotated
their D-term into the superpotential, we have a complex
electric FI coefficient $(e-i\xi)$.

Expanding the field $\p$ around its background value $\p_0$ as
$\p=\p_0+\vp$, we find
\eqn\acatp{ S= -Im\int d^4 x \int d^2 \t \bigg[
\tau(\p_0) (\ha W^\a W_\a + \vp \bD^2 \bar\vp + m\vp) + (e-i\xi)\vp
+ ...
\bigg]
\ , }
where $\tau (\p) = \cF''(\p)$ and
the neglected terms are higher-order in the fluctuations
(or independent of them).
As in \apt, the coupling is determined in terms of
the FI terms by the condition that terms linear in $\p$ cancel:
$\tau(\p_0)=-(e-i\xi)/m$.

Sending $\tau(\p_0)$ to infinity in a general complex direction sends
the mass of the chiral multiplet to infinity and decouples it. At the same
time, this produces the constraint we had before (see \cti). This
both gives the rationale behind and makes precise the relation between the
model of \apt\ and the supersymmetric BI action (expected on general
grounds in
\bgg).

More precisely, one is
expanding near the classical solution for which $\tau=\const$
and is to
integrate out the massive chiral multiplet $\vp$.
The leading (derivative-independent)
part of the resulting low-energy effective action
for the vector multiplet is then
the $N=1$ supersymmetric BI action.
One may drop the derivative terms for $\vp$
(as they would give derivatives of the vector field strength
which may be ignored at low energies)
and then solve for the scalar multiplet classically
(i.e. integrate it out including only tree diagrams).
This is similar to how the BI action
is derived in string theory by integrating out massive string modes
(see section 3.1 below).
Sending $\tau(\p_0) \to \infty$ corresponds to
decoupling the propagating massive
degrees of freedom of the chiral multiplet and
thus
effectively represents
the procedure of integrating out the
massive states.\foot{The above
short-cut $\tau\to \infty $ argument
should apply after a field redefinition
which effectively decouples the fluctuation field
allowing one to ignore correction terms in \acatp.
Simply dropping
all
massive mode contributions by sending their
mass to infinity would give
the Maxwell action instead of the BI action: one must
first make a field redefinition
that effectively accounts for the relevant coupling between
the vector and scalar fluctuations, and then ignore irrelevant
scalar multiplet couplings by sending its mass to infinity.}

We see the complete universality of the resulting BI action -- nothing
depends on the choice of $\cal F$ since $\tau$ appears linearly in the
action and plays the role of the Lagrange multiplier.
But the effective
action does depend on the parameters $m,e,\xi$
of the `microscopic' theory:
they appear in the overall (coupling constant)
coefficient, $\theta$-term and the fundamental scale of the BI action:
\eqn\effe{
S_{eff} = \int d^4 x \bigg( { m^2 \ov g^2} \big[
1- \sqrt {- \det (\eta_{ab} + m\inv F_{ab}) }\big] + \fo \theta F^{ab}
F^*_{ab} \bigg) \ , }
$$
g^{-2} \equiv { \xi\ov m} \ , \ \ \ \ \ \ \ \
\theta \equiv { e \ov m} \ . $$
The model \fif\ does not have a direct generalisation to the presence of
other matter hypermultiplets {\it minimally } coupled to the vector
multiplet \pap. This is not a problem in principle -- the action we are
interested in, such as \one, should contain non-minimal couplings.
Indeed, there should exist an $N=4$ supersymmetric 3-brane action
with 6 scalars coupled
to a $U(1)$ vector with non-linearly realised supersymmetry.
Given that the pure BI action (with no scalars) follows from
the model of \apt\
in the large mass limit, one should expect that such an action (where scalars
are coupled to a vector in a non-minimal way and are actually neutral with
respect to
$U(1)$) should also follow from some generalisation of the `dual D-term'
model.
\subsec{The Tensor Multiplet}
We now consider the $N=2$ tensor multiplet and find a Goldstone
multiplet expressed in terms of an $N=1$ real linear superfield. The
procedure we follow is entirely parallel to the vector case: we partially
break supersymmetry by choosing a particular background, and then
eliminate the `radial' fields by imposing a nilpotency constraint on the
fluctuations.
The $N=2$ tensor multiplet is described by a pure imaginary isotriplet of
scalar fields $\LL_{ab}$ $(a,b=1,2)$ satisfying \tens
\eqn\dlab{D_{\a(a}\LL_{bc)}=0\ , \ \ \ \ \ \ \ \ \LL_{ab}=-C_{ac}C_{bd}
\bar \LL^{cd}\ , }
or, equivalently,
\eqn\qlab{D_\a \LL_{11}=0\ ,\ \ Q_\a \LL_{11}=-2D_\a \LL_{12}
\ ,\ \ 2Q_\a \LL_{12}=-D_\a \LL_{22} \ ,\ \ Q_\a \LL_{22}=0\ .}
In just the same way as for the vector multiplet, we impose
the constraints:
\eqn\labcon{ Q^2 \langle\LL_{11}\rangle =
- QD \langle \LL_{12}\rangle = -1
\ , \ \ \ \ \ \ \ \ (\LL_{11}-\langle \LL_{11}\rangle )^2=0\ , }
\ie\ 
$\langle \LL_{11}\rangle = \ha\t_2^2$ and $\langle \LL_{12}\rangle =
-\ha\t_1\t_2$.
We define the
$N=1$ component fields $\p,G$ as follows:
\eqn\fig{\p = \LL_{22}|_{\t_2=0}\ ,\ \ \ \ \ G = -2 \LL_{12}|_{\t_2=0}\ ,\ \
\ \
\ \ \bp = -\LL_{11}|_{\t_2=0}\ .}
As for the vector multiplet, we apply $\bar Q^2$ to the nilpotency
constraint to find:
\eqn\tenseq{0=\bar Q^2(\ha\p^2)=\p \bar Q^2 \p +\ha\bar Q^\da\p\bar
Q_\da\p = \p -\p\bD^2\bp+\ha \bD^\da G \bD_\da G\ . }
This constraint can be solved straightforwardly to express $\p$ in terms of
the
linear superfield $G$ that describes the tensor multiplet,
leading to
the following Lagrangian for $G$
\eqn\aci{
L_G= - \ha G^2 + {\ha (D^\a GD_\a G)( \bD^\da G\bD_\da G) \ov
1 - \ha(\G^2 +\bar \G^{2}) +
\sqrt { 1 - ( \G^2 +\bar \G^{2}) -\qu (\G^2 - \bar \G^2 )^2 }}\ , }
where $$\G^2=(D^\a\bD^\da G)(D_\a\bD_\da G)\ , \ \ \ \ \ \ \ \
\bar\G^2=(\bD^\da D^\a G)(\bD_\da D_\a G) \ . $$
This agrees with the
results of \bggg; however, our derivation explains
why the final expression is unique.
\subsec{Duality and the Chiral Mutliplet}
The action \aci\ may be dualized to replace the real
linear superfield $G$ that describes the tensor multiplet by a chiral
superfield $\vp$ in the standard way: we relax the linearity constraint
on $G$, introduce a chiral Lagrange multiplier $\vp$ to re-impose the
constraint, and eliminate $G$. We start with
\eqn\acivp{
L_G'= L_G+(\vp+\bar\vp)G\ , }
and vary with respect to $G$. Unfortunately, the resulting equations are
difficult to solve in a closed form. Comparing directly to the bosonic
action, once can easily deduce the form that the action must take:
\eqn\acvp{
L_\vp =\vp\bar\vp+{\ha (D^\a \vp D_\a \vp)( \bD^\da \bar\vp\bD_\da \bar\vp) \ov
1 + A + (D^2\vp\bD^2\bar\vp)f
+\sqrt { (1 + A)^2 -B + (D^2\vp\bD^2\bar\vp)g }}\ , }
where $$A=\pa^\ada\vp\pa_\ada\bar\vp\ , \ \ \ \ \ \ \
B=(\pa^\ada\vp\pa_\ada\vp)
(\pa^\ada\bar\vp\pa_\ada\bar\vp)\ , $$
and $f$ and $g$ are unknown functions
of $A$, $B$ and $D^2\vp\bD^2\bar\vp$ that do not change the bosonic part
of the action.\foot{Actually, without loss of generality, we may drop
$g$ by a shift of $f$.}
In a subsequent article \gpr, we have verified
that the action \acvp\ is indeed dual to \aci, and have shown that the
functions $f,g$ are arbitrary, and may be chosen to vanish following
suitable redefinitions of $\vp$. We observe that the action \acvp\ has
manifest {\it translational symmetry}  and thus {\it is} indeed the action
representing  the 3-brane in 6 dimensions of ref.
\hulp. This contrasts with ref.\
\bg, where the two leading terms in the corresponding chiral multiplet
action were proposed. These terms are {\it not} invariant under a shift
of the bosonic part of the chiral superfield $\vp$ by a constant and thus
are not directly related to the supersymmetric form of the $D=6$ 3-brane
action. The relation, however, can be established by making a field
redefinition that eliminates non-translationally invariant terms in the
$O(\vp^2) + O(\vp^4)$ action of \bg.
\newsec{`Quadratic' form and duality transformations of Born-Infeld-Nambu
actions}
We now make some useful observations about the structure of the
bosonic actions \one,\two\ and their duality properties inspired in
part by the component expansions of our supersymmetric results.
\subsec{ $D=4$ Born-Infeld
Action in Terms of Two Auxiliary Complex Scalar Fields}
Let us start with the $D=4$ BI action and present it in a simple form
which is related to its supersymmetric
generalisation. Introducing an auxiliary field $V$, the BI Lagrangian
can be written as\foot{We use Minkowski signature $(-+++)$ so
that $\ep^{abcd}\ep_{abcd} =-1$, etc. Complex conjugation is
denoted by bar, Hodge duality by star ($F^{*ab}$), and
fields of dual theory by tilde ($\td A_a$).}
\eqn\bii{ L_{4} = - \sqrt{-\det (\eta_{ab} + F_{ab})} \
\to \ \ -\ha V \det (\eta_{ab} + F_{ab}) + \ha V\inv \ . }
Since in 4 dimensions
\eqn\foo{ -{\rm det}_4 (\eta_{ab} + F_{ab}) = 1 + \ha F_{ab}F^{ab} -
{\textstyle {1\ov 16} } (F_{ab}\F^{ab})^2 \ , \ \ \ \ \
\F^{ab}\equiv \hal\ep^{abcd} F_{cd} \ , }
we can put the action into a form {\it quadratic} in $F_{ab}$ by
introducing the second auxiliary field $U$ to `split' the quartic $(FF^*)^2$
term in
\foo\foot{Similar representations exist for BI actions in $D>4$
but are more complicated as they involve more auxiliary fields and
more field strength invariants (three in $D=6$, four in $D=8$, etc.).}
\eqn\biio{ L_{4} = \ha V + \ha V\inv + \ha V\inv U^2 + \fo V F_{ab}F^{ab} + \fo
U
F_{ab}\F^{ab}\ . }
Finally, we can eliminate the terms with $V^{-1}$ by
introducing a complex auxiliary scalar $a=a_1+ia_2, \
\bar a=a_1-ia_2$\
\eqn\biii{ L_{4} = \ha V (1 - \bar a a + \ha F_{ab}F^{ab} ) +
\ha U [i(a - \bar a) + \ha F_{ab}\F^{ab}] - \ha (a + \bar a)
\ . }
Shifting $a\to a + 1 $ and dropping the constant $-1$,
\biii\ can be rewritten as
\eqn\iii{ L_{4} = -\ha V (a + \bar a + \bar a a -
\ha F_{ab}F^{ab} ) + \ha U [i(a-\bar a)
+ \ha F_{ab}\F^{ab}] - \ha (a + \bar a ) \ , } or
\eqn\ciii{ L_{4} = - Im \bigg(\l \bigg[ a +
\ha \bar a a - \qu (F_{ab}F^{ab} +
i F_{ab} \F^{ab})\bigg] + i a\ \bigg) \ , }
$$
\l = \l_1 + i\l_2 \equiv U + i V \ . $$
Note that the constraint implied by $\l$ is solved by $a=a(F)$ with
$Im \ a (F) = \four F^{ab} F^*_{ab}$ and 
the real part
\eqn\solv{
Re\ a(F)\ = \ - 1 +
\sqrt { 1 + \ha F^2 - {\textstyle{1 \ov 16}} (F\F)^2 } \ , }
which is (up to sign) the
BI Lagrangian itself. This gives a natural `explanation' for the square
root structure of the BI action. One can thus view the BI action as
resulting from a peculiar action for 2 complex non-propagating
scalars $(\l,a)$ coupled non-minimally to a vector.

The bosonic part of the supersymmetric action \cti\ is
exactly the BI action represented in the form with two auxiliary
complex scalar fields \ciii , with the scalar fields $a$ and $\l$
in \ciii\ being
the corresponding scalar
components of the chiral superfields $\p$ and $\L$ in
\cti\ (note that $D^2 W^2 = - \fourth
F^{ab}F_{ab} - { i \ov 4} F^{ab}\F_{ab} + \ha {\rm D}^2$). It is thus
guaranteed that once we solve for $\L,\p$ the action \cti\ should
become the $N=1$ supersymmetric extension
\refs{\cecf,\bgg} of the BI action. It is clear from \cti\ that the bosonic
part of the
action is quadratic in the auxiliary field $ {\rm D}$ of $V$ so that
${\rm D}=0$ is always a solution; adding the FI term breaks $N=1$
supersymmetry giving a solution with ${\rm D}= \xi + O(\xi, F)$.

Shifting $\l$ by $i$
the action \ciii\
can be put into a form that does not contain terms linear in the fields
\eqn\cjj{
L_{4} = - \fo F_{ab}F^{ab} + \ha \bar a a - Im\bigg( \l \big[ a +
\ha \bar a a - \qu ( F_{ab}F^{ab}
+ i F_{ab} \F^{ab})\big] \bigg) \ . }
This may be viewed as a special case of the following
action for a vector coupled non-minimally to massive scalars
\eqn\acc{
L_4= - \fo F_{ab}F^{ab} -\ha (\del_a \vp_n)^2 - \ha m^2_n \vp_n^2 +
g_{nmk} \vp_n \vp_m \vp_k +
\vp_n (\a_n F_{ab}F^{ab} + \b_n F_{ab} \F^{ab}\big) \ . }
In the limit when the
masses of scalars are much larger than their gradients
so that the $(\del_a \vp_n)^2 $ terms may be ignored,
\acc\ reduces to \cjj\ with
the scalars $\vp_n$ being linear combinations of $\l_1,\l_2,a_1,a_2$
in \cjj.
This action may be viewed as a truncation of the cubic open string field
theory action
which reproduces the BI action as an effective action
upon integrating out (at the string tree level)
massive string modes $\vp_n$ \refs{\ft,\man}. The kinetic
term $(\del_a \vp_n)^2 $
may be dropped since it leads to the derivative
$\del_c F_{ab}$-dependent terms which, by
definition, are not included in the
leading part of the low-energy effective action action.

Let us note in passing that this
quadratic in $F_{ab}$ form of the BI action has an obvious
{\it nonabelian} generalisation.
As suggested by the
open string
field theory analogy, one replaces, \eg,
for $SO(N)$ case, 
$F_{ab}$ by an antisymmetric
matrix in the fundamental representation 
 and replaces the scalars $\vp_n$ (\ie , $\l$ and $a$) by {\it
symmetric} matrices. The result is
\eqn\cjjn{
(L_{4})_{nonab.} = \tr \bigg[ - \fo F_{ab}F^{ab} + \ha \bar a a -
Im \bigg( \l \big[ a + \ha \bar a a - \qu ( F_{ab}F^{ab}
+i F_{ab} \F^{ab})\big] \bigg) \bigg] \ . }
Integrating out the matrices 
$\l$ and $a$ one finds a nonabelian version of the $D=4$ BI
action in which all corrections to the standard YM $\tr F^2$ term
depend only on the
two symmetric matrices $ (F_{ab}F^{ab})^{pq}$
and $ (F_{ab}\F^{ab})^{pq}$.
It should be straightforward to write down the supersymmetric
generalisation of \cjjn. Like the symmetrized trace
action of \tst\ this non-abelian
action has two required
features:
the standard BI action as its abelian limit, and the single
trace structure. However, 
it is {\it different} from the
symmetrized trace
action (by some commutator terms)
already at the $F^4$ level. Indeed, the latter action 
\eqn\sstr{
{\rm Str}[I- \sqrt{-\det (\eta_{ab} + F_{ab})}
] = - \fo \tr FF + {\textstyle{ 1 \ov 32}} {\rm Str} [(FF)^2 + (FF^*)^2 ]
+ ...
\ }
contains the terms like
$\tr ( E_i E_j B_i B_j)$, $\tr ( E_i E_j E_i E_j)$ and
$\tr ( B_i B_j B_i B_j)$ (where
$E_i = F_{0i}, \ B_i =\ha \ep_{ijk} F_{jk}$)
which cannot be written in terms of $ (F_{ab}F^{ab})^{pq} = 2(B_iB_i - E_i
E_i)^{pq}$
and $ (F_{ab}\F^{ab})^{pq} = 4 (E_i B_i)^{pq}$ only
for generic internal symmetry group.
In contrast to \sstr\
it is not immediately clear how to generalise
the action \cjjn\ to dimensions higher than 4.

The BI action \bii,\foo\ is obviously invariant under $F_{ab} \to
F^*_{ab}$.
In addition, it is covariant under the vector-vector duality
transformation \refs{\gib,\tse}.
Since, in this form, the BI action is quadratic
in the vector field, it is very simple to demonstrate its covariance
under the duality.
Adding the Lagrange multiplier term $ \ha \td F^{*ab} F_{ab},$
where $\tF_{ab}$ is the strength of the dual vector field, 
and integrating out $F_{ab}$ we find that the dual action has the same
form as \ciii\ with ({cf.} \dal)\foot{The equations of motion derived
from the
vector terms in the action \ciii\ have the full
$SL(2,R)$ invariance ({cf.} \gaz):
$
\l \to { m \l + n \ov k \l + l } , $ $ \ \
F_{ab} \to (k U + l) F_{ab} + k V F^*_{ab} , $ $\ ml-nk=1.
$}
\eqn\dual{F_{ab} \to \tF_{ab} \ , \ \ \ \ \ \ \
\l \to - { \l\inv }\ , \ \ \ \ \ \ \ \ \ a \to - i \l a \ . }
As in the Maxwell theory case, the action \ciii\ is not invariant under this
duality. There exists, however, an equivalent
action containing one extra vector field variable
which is manifestly duality-symmetric (similar
duality-symmetric actions were constructed
in $D=2$ \refs{\tsd,\ss} and $D\geq 4$ \ss).
A systematic way of deriving such
duality-symmetric actions is explained
in the Appendix. Given an action (\eg, for a $D=4$ vector)
that depends only on the field strength, one
puts it in a special first-order form
by gauging the symmetry $A_a \to A_a + c_a, \ c_a=\const$ as follows.
One introduces a gauge field $V_{ab}$; then minimal coupling implies that
$F_{ab}$ is replaced by $F_{ab}+ V_{ab}$. One imposes
the constraint $dV=0$ with a Lagrange
multiplier (dual field) $\td A_a$
(\ie , one adds the term $\ha \td F^{*ab} V_{ab}$).
Choosing the `axial' gauge $V_{ij}=0$ ($i,j=1,2,3$)
and integrating out $V_{0i}$ then leads to an action for
$A_i$ and $\td A_i$ that is manifestly duality-invariant,
\ie , invariant under the interchange of $A_i$ and $\td A_i$.
The lack of manifest Lorentz invariance is not a problem, as it is merely
a consequence of a {\it noncovariant gauge choice}.

The `quadratic' form of the BI action \ciii\
makes it easy to obtain this duality symmetric version
of the action.
One finds ($I,J=1,2$; \ $i,j,k=1,2,3$)
\eqn\ddd{
\hat L_4(A,\td A) = - \ha \big[ \E^{I}_i {\cal L}_{IJ} \B^J_i +
\B^{I}_i ({\cal L}^T {\cal M L})_{IJ} \B^{J}_i \big]
-  Im \big[ \l (a + \ha \bar a a) + i a \big ] \ , }
$$ \E^I_i = \del_0 \A^I_i = (E_i, \td E_i) \ , \ \ \
\B^{I}_i = \ep_{ijk} \del_j \A^I_k = (B_i, \td B_i) \ , \ \ \ \
\A^I_i= ( A_i, \td A_i) \ ,
$$
\eqn\maat{ {\cal L} = \pmatrix{ 0 & 1 \cr
-1 & 0 \cr}\ , \ \ \ \ \ \ \
{\cal M} = { 1 \ov \l_2} \pmatrix{ 1 & \l_1 \cr
\l_1 & |\l|^2 \cr}\ . }
The vector field terms in this action are the same as in the
case of the standard
abelian vector theory coupled to scalars
\sss\
and are
invariant under the $SL(2,R)$ duality transformations
\eqn\duul{
\A_i \to \omega^T \A_i
\ , \ \ \ \ {\cal M} \to \omega^T {\cal M} \omega \ , \ \ \ \
\omega {\cal L} \omega^T = {\cal L}\ , \ \ \ \ \
\omega \in SL(2,R) \ . }
Because  the last term in \ddd\ is linear in $a$,
it can absorb any
variation, {\it i.e.}  we can choose its
transformation to extend the invariance to the complete action.

The `doubled' form of the BI action depending only on
$A_i,\td A_i$ can be obtained either by eliminating the scalar fields
from \ddd\ or, directly,
from the original action \bii,\foo\ by using
the procedure of gauging
($F_{ab} \to F_{ab} + V_{ab}$, \ $L_4 \to L_4 +
\ha \td F^{*ab} V_{ab}$), choosing the
`axial' gauge ($V_{ij}=0$) and integrating out $V_{0i}$
as explained above and in the Appendix.
The gauge-fixed action is found to be
\eqn\stee{\hat L_4 =- v_i\td B_i - \sqrt{ 1 - (E_i + v_i)^2 + B^2_i
- [(E_i+v_i) B_i]^2 } \ , \ \
\ \ \ \ v_i\equiv V_{0i} \ . }
Shifting $v_i\to v'_i -E_i $ and eliminating $v'_i$ from the action
by solving its equation of motion
gives
(after symmetrising the first term using integration by parts)
\eqn\finn{
\hat L_4 = \ha (E\cdot \td B - \td E\cdot B )
-
\sqrt{ 1 + B^2 +
\td B^2 + B^2 \td B^2 - (B\cdot \td B)^2 } \ , }
which can be represented also in the form (cf. \ddd)
\eqn\mani{
\hat L_4 = \ha \E^I_i \cL_{IJ}\B^J_i
-
\sqrt{ \det( \delta_{ij} + \B_i^I \B_j^I ) } \ . }
This is obviously invariant under the $O(2)$ duality
rotations in $(A, \td A)$ plane, in particular, under
$ A_i \leftrightarrow
- \td A_i$, \ie ,
\eqn\inin{
\cA \to \cL \cA \ , \ \ \ \
\E_i \to \cL \E_i \ , \ \ \ \
\B_i \to \cL \B_i\ . }
\subsec{Scalar-Tensor Duality and Born-Infeld-Nambu Actions}
Starting with the Dp-brane action like \one, \ie ,
\eqn\dde{
S= - \int d^{p+1} x \sqrt{- \det( g_{ab} + F_{ab})}\ , \ \ \ \ \ \ \
g_{ab} = \eta_{ab}
+ \del_a X^n \del_b X^n \ , }
and performing a vector duality transformation
by adding a Lagrange multiplier term $ \ha H^{*ab}F_{ab}$
and integrating out $F_{ab}$ one finishes (in $D\leq 5$)
with the same action with
$F_{ab} \to H_{ab}$ \tse.\foot{In $D=4$ this is true for an
arbitrary 4-d metric $g_{ab}$. In $D=5$ where $H^{* ab}
= {1\ov 6} \ep^{abcde} H_{cde}$ one finds the dual Lagrangian
in the form
$ \sqrt{- \det( g_{ab} + {g_{ac} g_{bd} \ov \sqrt g} H^{*cd})}$.}
Here we study what happens
if instead we dualise one of the scalars, \eg , $Y\equiv X^1$.
Since for general $D$
\eqn\dee{
\det(M_{ab} + P_a P_b ) = [ 1 + (M^{-1})^{ab}
P_a P_b ]\ \det M \ , }
where, in the present case $P_a = \del_a Y$ and ($s\not=1$)
\eqn\deer{
M_{ab} = \hat g_{ab}+ F_{ab} \ , \ \ \ \
\hat g_{ab}\equiv \eta_{ab}
+ \del_a X^s \del_b X^s \ , }
we can write the action in the form similar to \bii\foot{Note that
in contrast to the vector case, one auxiliary field suffices
to make the action quadratic in the scalar field derivatives.
}
\eqn\sii{
L_{D} =- \ha V [ 1 + (M^{-1})^{ab} \del_a Y \del_b Y]\
\det M + \ha V\inv \ . }
Replacing $\del_a Y $ by $P_a$ and
adding the Lagrange multiplier term $H^{*a} P_a, $ \
$(H^{*a} \equiv $ $ { 1 \ov p!} \ep^{a b_1 ...b_p} H_{b_1 ...b_p},
\ H_{b_1 ...b_p} \equiv p \del_{[b_1} \td Y_{b_2 ...b_p]}$)
one can integrate over $P_a$ to get\foot{We are grateful 
to S. Kuzenko for pointing out a mistake in this equation 
in the original version of this paper.}
\eqn\stii{
\td L_{D} = -\ha V \det M +
\ha V\inv \det M\inv\ \bar g_{ab} H^{*a }H^{*b } + \ha V\inv \ , }
$$ \bar g _{ab}  \equiv  [(M^{-1})^{(ab)}]^{-1} =
\hat{g}_{ab} -
F_{ac} \hat{g}^{cd} F_{db}
\ , $$ 
or, equivalently,
\eqn\ytr{
\td L_{D}=\ha \hat V ( \det M +
{ \bar g_{ab}} H^{*a } H^{*b} )
- \ha \hat V \inv \ , \ \ \ \ \ \ \
 \hat V \equiv V\inv \det M\inv \ . }
Using \dee\ this can be put back into the `determinant' form
\eqn\styii{
\td L_{D} = -\sqrt{- (\det M +
{ \bar g_{ab}} H^{*a} H^{*b})} \ . }
In $D=2$,
the action takes the same form when written in terms
of $\del_a \td Y$, \ $H^{*a}=\ep^{ab} \del_a \td Y$,
\ie , the $D=2$ Born-Infeld-Nambu action is invariant under the scalar-scalar
duality (note that
${\rm det}_2 M_{ab} = \det \hat g - \fourth (\ep^{ab} F_{ab})^2$)
\eqn\spec{
\td L_2 =- \sqrt{- \det (\eta_{ab} + \del_a X^s \del_b X^s +
\del_a \td Y \del_b \td Y + F_{ab}) } \ . 
}
In $D=3$, we get an action with
an extra vector $\td Y_a$ instead of the scalar $Y$
($H^{*a} = \ha \ep^{abc} \td F_{bc}, \ \hat g \equiv \det \hat g_{ab}$)
\eqn\spc{
\td L_3 = -\sqrt{-
[\ \hat g (1 + \ha \hat g^{ac} \hat g^{bd} F_{ab}F_{cd} )
+ \ha \bar  g_{ab} H^{*a} H^{*b }] } \ .
}
Note that dualising a scalar in $D=3$ we do get an action with {\it
two}
vectors but not simply
the usual $D=3$ BI-Nambu action
with $F_{ab} \to F_{ab} + \td F_{ab}$ (the determinant in such an
action would contain the
cross-term $F^{ab} \td F_{ab}$).
The inverse to this $D=3$ transformation, \ie ,
`vector $\to $ scalar'
duality (previously discussed in \refs{\town,\tse})
gives the membrane action with one extra scalar and no vectors.

In $D=4$, the scalar $Y$ is traded for the antisymmetric
tensor $B_{ab}\equiv \td Y_{ab}$
\eqn\pqc{
\td L_4
= - \sqrt{- [\det(\hat g_{ab} + F_{ab}) + \bar g_{ab} H^{*a} H^{*b}] }
\ , \ \ \ \ \  \ \ \  H^{*a}= {1\ov 3!}  \epsilon^{abcd} H_{bcd} \ . } 
These actions can be simplified
when $F_{ab}=0$ and $\hat g_{ab} =\eta_{ab} + \del_a X \del_b X$,
which, in particular, is the case of the
3-brane ($D=4$) in 6 dimensions \two.
For arbitrary dimension $D$,
$$\ (\eta_{ab} + Q_a Q_b )\inv = \eta_{ab} - {Q_a Q_b\ov 1 +Q^2} \ , $$
$$ -\det (\eta_{ab} + Q_a Q_b + P_a P_b )
= 1 + Q^2 + P^2 + Q^2 P^2 - (QP)^2 \ , $$
so that (here $ Q_a = \del_a X, \ P_a = \del_a Y$;
we use the flat Minkowski metric $\eta_{ab}$ to contract the indices)
\eqn\piqc{
L_D =- \sqrt{ 1 + (\del X)^2 + (\del Y)^2 +(\del X)^2 (\del Y)^2 -
(\del X\del Y)^2 } \ .
}
The dual action
\eqn\iqc{
\td L_D = - \sqrt{ 1 + \del_a X \del_a X -
H^{*a} H^{*a} - (H^{* }_a\del_a X )^2 } \ ,
}
can be expressed in terms of a complex vector
$\G_a$
\eqn\comp{
\td L_D =- \sqrt{ 1 + \ha ( \G_a \G_a + \bar \G_a \bar \G_a)
+ {{\textstyle{1\over 16}}} (\G_a \G_a - \bar \G_a \bar \G_a )^2 } \ , \ \
\ \ \
\G_a \equiv \del_a X + i H^*_a \ . }
In $D=4$, $H^{*a} = {1 \ov 6} \ep^{abcd} H_{bcd}$.
This form of the action can be compared with the bosonic
part of the tensor multiplet action \aci.

The scalar action \piqc\ may be put into a first-order form similar to
\biii\ in the case of BI action \bii.
Introducing $\vp = X + i Y,\ {\bar \vp} = X-iY$ and the real auxiliary
field $V$
we may replace \piqc\ by (cf. \bii)
\eqn\iqcr{
L_D = -\ha V\inv \big[ (1 + \ha \del \vp \del {\bar \vp})^2
- \four (\del \vp)^2 (\del {\bar \vp})^2 \big] - \ha V \ . }
Using one real ($\a$) and one complex ($\b$) auxiliary fields
we can put \iqcr\ into the form
quadratic in $\vp,{\bar \vp}$
\eqn\uuq{
L_D=- \ha V(1 + \bar \b \b - \a^2) + \a( 1 + \ha \del \vp \del {\bar \vp})
+ \fo \b (\del {\bar \vp})^2 + \fo \bar \b (\del \vp)^2 \ . }
The field $V$ thus plays the role of a Lagrange multiplier
which restricts $\a,\b$ to a 2-dimensional
hyperboloid.
This form of the action may be useful
for trying to construct a $N=1$ supersymmetric extension
of \piqc\ based on chiral multiplet.
Let us note also that \piqc\ can be also written in the form
\eqn\anou{
L_D = - \sqrt{ 1 + \del \vp \del {\bar \vp}
+ \four (\del \vp \del {\bar \vp})^2
- \four (\del \vp)^2 (\del {\bar \vp})^2 } }
$$= - 1 - \ha \del \vp \del {\bar \vp}
+ { \four (\del \vp)^2 (\del {\bar \vp})^2 \ov
1 + \ha \del \vp \del {\bar \vp} +
\sqrt{ (1 + \ha \del \vp \del {\bar \vp})^2
- \four (\del \vp)^2 (\del {\bar \vp})^2 } } \ , $$
which can be compared to the supersymmetric action \acvp.
\newsec{Conclusions }
There exists a remarkable connection between
(i) partial supersymmetry breaking, (ii) nonlinear realisations of
extended supersymmetry, (iii) BPS solitons , and (iv)
nonlinear Born-Infeld-Nambu type actions.
We have shown that the connection between
partial breaking of supersymmetry and nonlinear actions
is not accidental and has to do with constraints that
lead directly to nonlinear actions of BI type.
We believe that this {\it constrained superfield approach}
is the simplest
and most transparent way of deriving and using these actions.

$N> 1 $ susy can be partially
broken either by a non-translationally invariant
background (soliton) in a second-derivative higher-dimensional theory
or by a translationally invariant vacuum
in a nonrenormalisable
theory in four dimensions containing non-minimal interactions.
We have seen how to determine the resulting actions in a
model-independent way.

Inspired by our supersymmetric results, we have found a particularly
simple way of demonstrating the self-duality of the BI action. This led
us to a discussion of the general form of scalar-tensor duality in
$D=4$ actions, and duality in other dimensions.

We found a closed form of the action for the tensor multiplet
which after duality becomes the full nonlinear action of the $D=6$
3-brane. A direct derivation of this action
using the methods of \hulp\ would be complicated,
whereas our approach gives it in a universal way.

Unfortunately, we have not found the $D=4$, $N=4$ supersymmetric extension
(with $N=2$
supersymmetry realized nonlinearly)
of the BI
action.\foot{The $N=2$ supersymmetric form of the BI action was recently
considered in \ket\ (see also \refs{\lamb,\ivano} in connection with $N=2$ actions). However, the action proposed there has higher
derivative terms
at the component level, and is not invariant under constant shifts of the
physical scalar components.}
To do this, we would need to couple the $N=1$ vector and chiral (or tensor)
multiplets.\foot{It may be of some use to note that BI action in $D=5$
is dual to the antisymmetric tensor
action for $H_{mnk}$, \ie,
$\sqrt {- \det {( \eta_{mn} + H^*_{mn}) }}$ \tse;
upon dimensional reduction to $D=4$ it becomes
an action for a vector $B_{a5}$ and a {\it tensor} $B_{ab}$
instead of a vector and a scalar as before duality in $D=5$. That suggests
a possibility of getting a manifestly supersymmetric action for a vector
multiplet
and a tensor multiplet by starting with a duality-rotated action for a vector
multiplet in $D=5$.}
 In \pap\ it was noted that there is no obvious generalisation
of the
model of \apt\ to the case of an $N=2$ vector multiplet coupled to a charged
hypermultiplet. This should not be a problem since we need a non-minimal
coupling -- as is evident from our results, scalars and vectors
should couple non-minimally via field strengths in BI actions.
So some version of constrained superfield approach should work.

It is not clear if the above constrained approach
can be generalised to the nonabelian case, where $\p$
in \cti\ will be in the adjoint representation.

\bigskip
\noindent{\bf Acknowledgements}
\smallskip
We would like to thank I.\ Antoniadis,
W.\ Siegel, F.\ Gonzalez-Rey, I.Y.\ Park and P.\ van Nieuwenhuizen
for useful discussions of
related questions.
The work of M.R.\ was supported in part by NSF grant No. PHY 97221101.
A.A.T.\ is grateful to the Institute of Theoretical Physics of
SUNY at Stony Brook
for hospitality during his stay in
April 1997 while most of this work was completed
and acknowledges also the support
of PPARC and the European
Commission TMR programme grant ERBFMRX-CT96-0045.
\vfill\eject
\noindent{\bf Appendix: Duality-symmetric actions -- the `gauging'
approach}
{\global\meqno=1\global\subsecno=0}
\smallskip
One particular definition of a duality relation between two quantum
field theories in $D$ dimensions is that it gives a map between
correlators of certain operators in one theory and correlators of
corresponding subset of operators in the dual theory.\foot{The dual theories may not be completely equivalent, \eg , in $S$-matrix
sense: only certain types of observables may be related.} At the level
of the partition function or the generating functional for the special
operators examples of such duality relations
can be understood as a series of formal
transformations of gaussian path integrals \ftt, but {\it not}
as a local change of variables in the original action.

At the same, it {\it is}
possible to define manifestly duality invariant actions
where duality becomes
just a local field transformation accompanied by
a transformation of the external coupling parameters
(\eg, $R\to \a'/R$). To achieve this
one doubles the number of field variables by introducing the dual
fields on the same footing as the original fields. This
was suggested in the context of scalar $D=2$ theories
in \tsd\ and generalized to the heterotic
string type $D=2$ actions in \ss\ and to
$D=4$ abelian vector actions in \sss\ (see also \dese\ for earlier
work).\foot{Such actions
(which are of first order in time derivatives)
can be also interpreted as phase space actions
(\ie, as original actions expressed in terms of phase space variables)
with the dual fields playing the role of the integrated canonical momenta;
for a related canonical approach to duality
see \canon.}
The price for having duality as a symmetry of the action
is the lack of manifest Lorentz invariance (the off-shell
Lorentz
invariance appears only after one integrates
out one of the dual fields recovering the original or dual
Lorentz-invariant actions, see also \refs{\jac,\teit}).

In what follows we demonstrate
that such non-Lorentz invariant `doubled' actions are, in fact, gauge
fixed versions of duality {\it and}\ Lorentz invariant `extended' actions
containing extra degrees of freedom. An application of this observation
to the nontrivial example of the bosonic Born-Infeld action
was considered above in section 3.1 (see \ddd,\mani).

These `extended' actions
are closely related to first-order
actions (see, \eg, \refs{\ftt,\busc,\rocver,\wit})
usually discussed in the context of path
integral demonstration of duality.
Given an action that depends only on a field strength $F=dA$,
one can treat $F$ as an independent field by adding a constraint
that imposes the Bianchi identity, thus relating it to the original
field. An equivalent first-order
action is obtained by gauging
the symmetry $A\to A + c,$ where $c$ is a constant
\rocver. The corresponding gauge field will be denoted by $V$. Its field
strength is set equal to zero by a Lagrange multiplier that plays the role
of the dual field $\td A$. Gauge fixing the original field and
integrating out the auxiliary field $V$ gives the dual action. If,
instead, one chooses a non-Lorentz-covariant `axial' gauge on $V$ and
integrates out the remaining components of $V$, one obtains precisely
the duality symmetric `doubled' action for $A$ and $\td A$.

An advantage of
considering the generalised actions
depending on $A,\td A$ {\it and } $V$ is that in addition to be duality
invariant (\ie, invariant
under $A\leftrightarrow \td A$ and a gauge transformation
of $V$), it is also Lorentz invariant. Thus it may be used as a starting
point for constructing manifestly Lorentz and
duality invariant world sheet and space-time
actions in string theory. The additional
variables $V$ may eventually find their
place in a  more fundamental formulation of the theory.

To illustrate this procedure
let us start with an action for a
scalar in $D=2$ or a vector in
$D=4$ that depends on the field $A$
only through the field
strength, $\ S= \int d^D x\ L (F) ,
\ F= dA, $
where $L$ may depend also on other fields.
For example, in $D=2$,
$L_2 = - \ha F_a F^a , \ F_a = \del_a A$. In $D=4$ one may consider
an action
for several vector fields $A^p_a $ coupled to scalar fields $\vp$:
$L_4(A,\vp) =
G_{pq} (\vp) F^p_{a b} F^{qab } +
B_{pq} (\vp) F^p_{ab} F^{*qab}
+ J^{ab}_p (\vp) F^p_{ab} + L'_4 (\vp)$, where
$a,b=0,1,2,3$.
The
`gauged' forms of the extended
`first-order' Lagrangians are\foot{For simplicity here
we shall consider the case of a single scalar ($D=2$) or vector ($D=4$).
Various possible generalisations
(\eg, to several scalar or vector fields,
curved $D$-dimensional space-time,
other $p=\ha (D-2)$ -form dualities, etc.) are straightforward.}
\eqn\saf{\hat L_2(A,\td A,V) = L_2 (V_a + \del_a A ) + \ep^{ab} \del_b \td
A \ V_a \ , }
\eqn\vee{\hat L_4(A,\td A,V) = L_4 (V_{ab} + \del_{a }A_{b}- \del_b A_a )
+ \ha \ep^{abcd} \del_a \td A_b V_{cd } \ . }
The corresponding actions are
invariant under the gauge transformations
\eqn\gau{ A'=A + c \ , \ \ \ \ \ V' = V - d c \ , }
as well as under the duality transformations that
interchange $A$ and $\td A$ and act on $V$ in a nonlocal way.

Integration over $\td A$ in \saf,\vee\ gives back the original
actions $\int L_D(dA)$
(after a redefinition of $A$ or gauge-fixing the
remaining longitudinal part of $V$ to zero).
Gauge-fixing $A=0$ and integrating out $V$ gives
the dual Lagrangian $\td L_D (\td A)$. Classically,
$V$ is eliminated by solving its equations of motion.
The resulting classically equivalent Lagrangian is local
provided $L_D$ is an {\it algebraic} function of $F=dA$,
as, \eg, in the case of the Born-Infeld action
discussed in \refs{\gib,\tse} and above.

Fixing instead the `axial' gauge on $V$, namely,
$V_1=0$ in $D=2$ and $\ep^{ijk} V_{jk}=0$ ($i,j=1,2,3$) in $D=4$
and integrating
out the remaining gauge field components (\ie, $V_0$ and $V_{0i}$)
we get from \saf,\vee\ in the simplest $L_2=-\ha F^2_a, \ L_4 = -\fourth
F^2_{ab}$ cases
(we use the freedom to add a total
derivative to put the Lagrangians in a more symmetric form)
\eqn\saff{\hat L_2 (A,\td A) =
- \ha \big[-\del_0 A \del_1 \td A -\del_0 \td A \del_1 A
+ (\del_1 A)^2 + (\del_1 \td A)^2\big] \ , }
\eqn\veel{\hat L_4(A,\td A) = -\four \big[ - F_{0k}\ep^{ijk}\td F_{ij}
+ \td F_{0k} \ep^{ijk} F_{ij} + F_{ij} F^{ij} + \td F_{ij} \td F^{ij}\big] }
$$ = -\ha \big[ - \del_0 A_k \ep^{ijk} \del_i \td A_j
+ \del_0 \td A_k\ep^{ijk} \del_i A_j + (\ep^{ijk} \del_{i} A_{j})^2
+ (\ep^{ijk}\del_{i} \td A_{j})^2 \big] \ , $$
or, in a compact symbolic form,
\eqn\gene{ L_D({\cal A} ) = \ha( \del_0 {\cal A} \ {\cal L } \
\del {\cal A}
- \del {\cal A} \ {\cal M} \ \del {\cal A}) \ , \ \ \ \
\ \ \ \ {\cal A} = (A, \td A)\ , }
$$ {\cal L} = \pmatrix{ 0 & I \cr
(-I)^{{D-2\ov 2}} & 0 \cr}\ , \ \ \ \ \ \ \ \ \ \
{\cal M} = \pmatrix{ I & 0 \cr
0 & I \cr}\ . \ \ $$
The actions \saff\ \tsd\ and \veel\ \sss\ are clearly
invariant under the discrete duality transformations
$\A \to {\cal L} \A$, \ie,
\eqn\dua{
D=2: \ \ A \to \td A\ , \ \ \ \td A \to A\ ;
\ \ \ \ \ \ \ D=4: \ \ A_i \to - \td A_i \ , \ \ \ \td A_i \to - A_i\ . }
In more general cases they are invariant under
\dua\
accompanied by a redefinition of the `external' coupling constants
(scalar fields)
which parametrise the matrix $\cal M$.
Similar action is found in the BI case (see \ddd,\finn).

The action \vee\ is the `master action' which unifies various
other actions with less number of field variables but the same number of
physical degrees of freedom. The `axial' gauge choice we used
is not the only possible one.
For example, one may choose $V_{ab}$ in the form
\eqn\gauu{ V_{ab} = u_a v_b -u_b v_a \ , \ \ \ \ \ \ u_a v_a=0 \ , \ \ \ u_a=
\del_a\psi \ , }
where $v_a$ is an arbitrary vector field
which is supposed to be integrated out while $\psi$ is kept along with
$A_i,\td A_i$.
In the case when $\psi=t$, \ie,
$u_a=(1,0,0,0)$, we are back to the axial gauge, $V_{ij}=0$,
$\ V_{0i} = v_i$, and integrating out $v_i$ leads to \veel.
Keeping $\psi$ arbitrary
one obtains the
`covariantized' form of the action \veel\
suggested in \pas\ (which does not contain
extra dynamical degrees of freedom
present in the earlier
proposal of \bra)
\eqn\pasti{
L_4(A,\td A,u) =
- \four F^{ab}F_{ab} +
\four (F^{ab} - \td F^{*ab}) (F_{ac} - \td F^*_{ac}) {u_b u^c\ov u^2}
\ }
$$ = - {\textstyle {1\ov 8}} ( F^{ab}F_{ab} +\td F^{ab}\td F_{ab} )
+ {\textstyle {1\ov 8}} \bigg[(F^{ab} - \td F^{*ab}) (F_{ac} - \td F^*_{ac})
+ (\td F^{ab} - F^{*ab}) (\td F_{ac} - F^*_{ac}) \bigg]
{u_b u^c\ov u^2} \ .
$$
The residual gauge invariance
of this action allowing one
to choose a gauge $u_a =\delta^0_a$ \pas\
(in which \pasti\ reduces back to \veel)
is now understood as being just a remaining part
of the original gauge invariance \gau\ of the
`master action' \vee.

We would like to emphasize that it is the `gauged'
first-order action
\vee\ depending on $A_a, \td A_a$ {\it and} $V_{ab}$
that is the {\it genuine Lorentz-covariant
action behind both the duality symmetric `doubled' action \veel\ of
\sss\
and the action with an extra vector $u_a$ of \pas}.

Let us also note that to obtain the corresponding actions
describing {\it self-dual} fields
one just sets $\A=\cL \A$
in the above `doubled' actions.
In particular,
this leads to a simple `self-dual vector' action in the BI case:
as follows from \finn,
\eqn\bis{
L_4^{(+)} = E_i B_i - \sqrt{ 1 + 2 B_i B_i } \ , }
which generalises the action of the self-dual
analogue $ (E_i B_i - B_i B_i) $ \teit\ of the Maxwell action.

The approach discussed above can be readily applied
to $D=6$ antisymmetric tensor theories, \eg, to the 5-brane theory
containing 2-tensor with self-dual field strength (see, \eg, \oth).
The `master' action in this case depends on $B_{ab}, \td B_{ab}$ and
$V_{abc}$; self-duality is imposed by identifying the spatial components of
$B_{ab}$ and $
\td B_{ab}$ after (partial) integrating out $V_{abc}$.

An alternative but equivalent procedure
is to start with the (in general, nonpolynomial)
action $\int L(dB)$
for a p-form field (\eg, $p=2, D=6$), write down the corresponding
Lagrangian in terms of the phase space variables, \ie,
the fields
$B_{ij}$ and $p^{ij} = \del L/\del (\del_0 B_{ij})$,
and replace the momentum $ p^{ij}$ by a new field $\td B_{ij}$,
\ $ p^{ij} = \ha \ep^{ijki'j'} \del_k \td B_{i'j'}$. The result is
the duality-symmetric Lagrangian
which generalises \gene\ to the case of an arbitrary nonquadratic
function $L(dB)$, \ie, \
$\hat L = \ep_5 \del_0 B \td B - L( H_{ijk}, H_{0ij} (p(\td B))),
$ where $H_{abc} =3\del_{[a } B_{bc]}$.
To obtain the action for a self-dual field
one sets $B=\td B$.
The `Lorentz-covariant' version of the resulting action
(generalising that of \pasi\ in the case of the quadratic action \gene)
is found by repeating
the above steps in the `covariant canonical formalism'
set-up, \ie, with $dt\to \del_a \psi dx^a$, etc.,
thus introducing the dependence on $u_a = \del_a \psi$ as in \pasti.
\vfill\eject
\listrefs
\end